\input phyzzx
%
%
%
\newcount\lemnumber   \lemnumber=0
\newcount\thnumber   \thnumber=0
\newcount\conumber   \conumber=0

\def\myeq{{\rm \chapterlabel\the\equanumber}}

\def\Lemma{\par\noindent\global\advance\lemnumber by 1
           {\bf Lemma\ (\chapterlabel\the\lemnumber)}}
\def\Corollary{\par\noindent\global\advance\conumber by 1
           {\bf Corollary\ (\chapterlabel\the\conumber)}}
\def\Theorem{\par\noindent\global\advance\thnumber by 1
           {\bf Theorem\ (\chapterlabel\the\thnumber)}}

%
%
\def\e{\adveq\eqno{\rm (\chapterlabel\the\equanumber)}}
\def\adveq{\global\advance\equanumber by 1}


%
%
\font\tensl=cmsl10
\font\tenss=cmssq8 scaled\magstep1
\outer\def\quote{
   \begingroup\bigskip\vfill
   \def\endquote{\endgroup\eject}
    \def\par{\ifhmode\/\endgraf\fi}\obeylines
    \tenrm \let\tt=\twelvett
    \baselineskip=10pt \interlinepenalty=1000
    \leftskip=0pt plus 60pc minus \parindent \parfillskip=0pt
     \let\rm=\tenss \let\sl=\tensl \everypar{\sl}}
\def\from#1(#2){\smallskip\noindent\rm--- #1\unskip\enspace(#2)\bigskip}

\def\WIS{\address{Department of Physics\break
      Weizmann Institute of Science\break
      Rehovot 76100, Israel}}

\def\r#1{$\lb \rm#1 \rb$}

%
%
\def\rarrow{\rightarrow}

\def\semidirect{\mathrel{\raise0.04cm\hbox{${\scriptscriptstyle |\!}$
\hskip-0.175cm}\times}}

\def\mod{\mathop{\rm mod}\nolimits}

\def\ref#1{$^{#1}$}

\def\twidle{\tilde}

\def\osum{\mathop\oplus\limits}

\def\half{{1\over2}}
\def\lb{\lbrack}
\def\rb{\rbrack}

\def\rank{\mathop{\rm rank}\nolimits}

\def\diam{{\hbox{\hskip-0.02in
\raise-0.126in\hbox{$\displaystyle\bigvee$}\hskip-0.241in
\raise0.099in\hbox{ $\displaystyle{\bigwedge}$}}}}

%


\def\sqr#1#2{{\vcenter{\hrule height.#2pt
      \hbox{\vrule width.#2pt height#1pt \kern#1pt
        \vrule width.#2pt}
      \hrule height.#2pt}}}

\def\underwig#1{	
	\setbox0=\hbox{\rm \strut}
	\hbox to 0pt{$#1$\hss} \lower \ht0 \hbox{\rm \char'176}}

\def\bunderwig#1{	
	\setbox0=\hbox{\rm \strut}
	\hbox to 1.5pt{$#1$\hss} \lower 12.8pt
	\hbox{\seventeenrm \char'176}\hbox to 2pt{\hfil}}

\def\gcd{\mathop{\rm gcd}\nolimits}

\Pubnum={}
\pubtype={}
\date{May, 1999}
\titlepage
\title{Realizations of Pseudo Bosonic Theories with Non--Diagonal Automorphisms}
\author{Ernest Baver, Doron Gepner and Umut G\"ursoy}
\WIS
\abstract
Pseudo conformal field theories are theories with the same fusion rules, but
different modular matrix as some conventional field theory. One of the authors
defined these and conjecture that, for bosonic systems, they can all be
realized by some actual RCFT, which is that of free bosons.
We complete the proof here by treating the non diagonal automorphism case.
It is shown that for characteristic $p\neq2$, they are equivalent to
a diagonal case, fully classified in our previous publication. For $p=2^n$
we realize the non diagonal cases, establishing this theorem.
\endpage
Rational conformal field theories in two dimensions have been the subject
of intense investigation in recent years. This is due to the fact that they
afford a tractable environment in which to test quantum field theory, along
with their cardinal role in string theory and condensed matter systems.
Several families of such theories were studied, for the most part stemming
from free bosons or the WZW--Sugawara affine construction.

In ref. \REF\Found{D. Gepner, Foundation of rational quantum field theory,
Caltech preprint, Nov. 92, Hepth 9211100}\r\Found, it was shown that a much
larger family of RCFT may exist, where one takes the same fusion rules as
the known ones, but with a different modular matrix. The latter theories are
termed pseudo conformal field theories.
The general philosophy is
that these new conformal data correspond to full fledged RCFT, which for the most
part, are yet to be explored. The simplest case of this is the bosonic and pseudo
bosonic systems, where we conjectured that all pseudo bosonic systems are
equivalent to some other conventional bosonic system.
In ref. \REF\BGG{E.
Baver, D. Gepner and U. G\"ursoy, On conformal field theories at fractional
levels, Weizmann preprint, Nov. 1998}\r\BGG, we proved the diagonal case.
Our purpose here is to find realizations for the non--diagonal case, thus completing
this theorem. In passing, we note that pseudo affine systems were treated
in ref. \REF\New{D. Gepner, On new conformal field theories with affine
fusion rules, Weizmann preprint, Jan. 1999}\r\New, where many new families
of theories were found, but much remains to be done.

The conventional bosonic theories are defined as a vector of free bosons,
$\vec \phi$ propagating on some lattice $M$. The primary fields are labeled by
elements of $M^*/M$, i.e., the dual lattice modulo the lattice. The `fusion
rules' are the way two fields fuse in the OPE,
$$[a]\cdot [b]=[a+b],\e$$
i.e., simple addition modulo $M$. Another important data is the modular matrix, $S$,
which implements $\tau\rarrow-1/\tau$, where $\tau$ is the modulus of
the torus. A relation observed by Verlinde \REF\Ver{E. Verlinde, Nucl. Phys. B300 (1988)
360.}\r\Ver\
relates the fusion rules and the modular
matrix,
$$f^\nu_{\lambda,\sigma}={\sum_a S_{\lambda,a} S_{\sigma,a} S_{\nu,a}^\dagger
\over S_{0,a}},\e$$
where $f^\nu_{\lambda,\sigma}$ is the fusion algebra structure constant.
Now, the point made in ref \r\Found\ is that taking fixed fusion rules and
solving for the modular matrix $S$, there are numerous solutions for any given fusion
rules, other then the known ones. These were termed pseudo conformal field
theories.

For the bosonic systems, let $G=M^*/M$ be the abelian group. The fusion rules
are the group algebra over $G$. For each such $G$ we can define some scalar
product, rather arbitrarily, which is a bilinear form of the group,
$g(\lambda,\mu)=\lambda\cdot\mu$. Now, the general solution to Verlinde eq.
(2), can be written as
$$S_{\lambda,\mu}=\exp[-2\pi i\lambda\cdot h(\mu)],\e$$
where $h$ is any symmetric automorphism of $G$, $h(\lambda)\cdot\mu=\lambda
\cdot h(\mu)$. This was proved in ref. \r\BGG.
The rest of the conformal data is given by,
$$\Delta_\lambda={\lambda h(\lambda)\over2}\mod Z,\e$$
$$e^{\pi ic/4}=|G|^{-\half}\sum_{\lambda\in M^*/M}  e^{\pi i\lambda\cdot h(\lambda)},\e$$
where $\Delta_\lambda$ is the dimension of $[\lambda]$, and $c$ is the central
charge of the theory (defined modulo $8$).

Now, if $G$ and $H$ are isomorphic groups, $G\approx H$, evidently they will
give precisely the same solutions to eq. (2). Thus, we may use any lattice
we wish, if it has the same group, $G\approx M^*/M$. Here, we invoke the basic
theorem of abelian groups, which says that any finite abelian group is
isomorphic to
$$G\approx \osum_i Z_{p_i^{n_i}},\e$$
where the $p_i$'s are some primes and $n_i$ some integers. Thus, it is enough
to consider the case of
$$N=\osum_i SU(p_i^{n_i})_1,\e$$
which has the same fusion rules, $G\approx N^*/N$.

Now, how does the general automorphism of $N$ looks like? We can choose a basis
for the lattice $N$ as a vector in each of the $SU(p_i^{n_i})$,
$$\lambda\cdot\mu=\sum_i \pi_i(\lambda)\pi_i(\mu),\e$$
where $\pi_i(\lambda)$ is the projection on the $i$th summand.
We can describe $h$, the automorphism as a matrix, $h_{ij}=\lambda_i \cdot
h(\lambda_j)$, where $\lambda_i$ is a generator of $SU(p_i^{n_i})$.
The matrix $h_{ij}$ is integral and symmetric. If $p_i\neq p_j$ then
$h_{ij}=0$. If $p_i=p_j=p$ then $h_{ij}p^{n_i}=0\mod p^{n_j}$
and $\gcd(\det(h),p)=1$. Any matrix
$h_{ij}$ obeying these condition gives rise to a symmetric automorphism,
and vice versa.

Now if $h_{ij}$ is a diagonal matrix we term this case regular. This case
was realized completely in ref \r\BGG. Our interest here is in the non--regular
case. Evidently we can limit ourself to $p_i=p_j=p$ for all $i$ and $j$, and
we can deal with the question prime by prime.

We wish to realize all the non--regular cases. What would be
a realization? Let $N_h$ be our conformal data, where $h$ is some automorphism.
We wish to find another lattice $M$ such that the RCFT, $M_1$, has $N_h$
as the conformal data. This is equivalent to finding an isomorphism $q$,
$$q:N^*/N\rarrow M^*/M,\e$$
such that
$$\lambda\cdot h(\lambda)/2=q(\lambda)^2/2\mod Z.\e$$
Our aim here is to realize all the non--regular cases, thus completing the theorem
mentioned in the introduction.

Note, that not all the matrices $h_{ij}$ give rise to a different theory.
As noted in ref. \r\BGG, we are always free to transform the primary fields
by some arbitrary matrix $B$, $\gcd(\det(B),p)=1$,
$$\lambda\rarrow B\lambda,\e$$
which is equivalent to changing the automorphism $h$ by a similarity transformation,
$$h\rarrow B^t h B.\e$$
Thus we need only realize one element from any of the similarity classes.

Let us, thus, turn to the realization of the non--regular cases. We need to
distinguish two possibilities: 1) $p\neq2$, 2) $p=2$. So, let us first
assume that $p\neq2$. Our claim is that by a similarity transformation any
non--diagonal automorphism can be seen to be equivalent to a diagonal one.

It is convenient to first concentrate on a rank $2$ automorphism $h$,
$$h_{ij}=\pmatrix{a&b\cr b& c\cr}.\e$$
and assume that $n_i=1$. This is the case where $G$ is actually a field.
Now, choose $B=\pmatrix{1&n\cr0&1\cr}$. Then, according to eq. (12), $h$ is
equivalent to
$$\pmatrix{1&0\cr n&1\cr} \pmatrix{a&b\cr b&c\cr}\pmatrix{1&n\cr0&1\cr}=
\pmatrix{*& an+b \cr*&*\cr}.\e$$
Now, if $a\neq0$ we may choose $n=-ba^{-1}\mod p$, which
exist since $p$ is prime. This means that for $a\neq0$ we may diagonalize the
automorphism.

If $a=0$ then we choose $B=\pmatrix{\alpha&\beta\cr\gamma&\delta}$ and
now $h=\pmatrix{0&b\cr b&0\cr}$ becomes
$$\pmatrix{\alpha&\gamma\cr\beta&\delta} \pmatrix{0&b\cr b&0\cr}
\pmatrix{\alpha&\beta\cr\gamma&\delta}=\pmatrix{*&b(\alpha\delta+\gamma\beta)\cr*&*\cr}.\e$$
Thus, we need to solve the equation,
$$\alpha\delta+\gamma\beta=0,\e$$
which always has solutions for $p\neq2$. (For $p=2$ it is equal to the determinant
which cannot vanish.) Say, we take $\alpha=\delta=\gamma=-\beta=1$.
It follows that $h$ can be assumed to be diagonal or a regular case.

Now for a prime power, $n_i>1$, the proof is the same. We have to separate
all the cases where $a=0\mod p$, but the proof works along the same lines.
We omit the details for brevity.

The same proof actually works for any rank of $h$. Say, $h$ is an $n\times n$
matrix. Then we diagonalize each $h_{ij}\neq0$, where $i\neq j$, in turn,
using a matrix with the only
non-diagonal entry $B_{ij}$. This is, in fact, very similar to the classical
diagonalization of symmetric matrices by similarity steps in
elementary algebra.

We conclude that for $p\neq2$ all automorphisms are indeed regular and thus
can be realized by the results of ref. \r\BGG.

For $p=2$ only one step in the proof fails. This is the second step, where
we assumed $a=0\mod p$. Thus if $a$ or $c$ are odd, it is still equivalent
to a diagonal automorphism. It is thus left to deal with the cases
$a=c=0\mod2$. We would like to know which of these are equivalent under field
transformation. For $\rank(h)=2$, the answer that we found is as follows.
There are for any $n_i$ exactly (we assumed that $n_i=n_j$;
this is due to the fact that for $n_i\neq n_j$, only diagonal
automorphisms exist, as can be seen by the fact that 
all the elements of the matrix $h$ must be even.)
two inequivalent non--regular automorphisms, which are
$$h=\pmatrix{0&1\cr1&0},\qquad \pmatrix{2&1\cr 1&2\cr}.\e$$
An automorphism equivalent to the first matrix, we call type
(I), and the second matrix type (II). Thus, all the automorphisms are diagonal, except for these two exceptional
types.
This we found by running a Mathematica program for $p=2^n$ up to $n=4$.

Subsequent to that we found and proved the following two theorems:

\Theorem:
The group of rank $m$ matrices modulo $p^n$ where $p$ is any prime, with determinant equal to $1\mod p$,
is generated by the following matrices,
$$A^{12}_{i,j}=\delta_{i,j}+\delta_{i,1}\delta_{j,2},$$
$$P^{1,a}_{i,j}=\delta_{i,j}-\delta_{i,1}\delta_{j,1}-\delta_{i,a}\delta_{j,a}+
\delta_{i,1}\delta_{j,a}+\delta_{i,a}\delta_{j,1},$$
which are the generators of permutations, and 
$$M^b_{i,j}=b\delta_{i,1}\delta_{j,1}+\delta_{i,j},$$
for $b=p^s\mod p^n$, and $s=1,2,\ldots,n-1$. Here, $i$ and $j$ label the entries of the matrices.

The proof of this theorem is outlined below. The second theorem is

\Theorem:
All the rank $m$ symmetric matrices over $2^n$, modulo the similarity transformation eq. (12), can be written in $2\times 2$ or $1\times 1$ block diagonal form, where the $2\times2$ matrices are of the three types mentioned above: diagonal, type (I) or type (II), eq. (17).
Let us now sketch the proof of these two theorems. Let $M_n$ be the 
group of such rank $m$ matrices modulo $p^n$. We have a natural map:
$$h:M_{n}\rarrow M_{n-1},\e$$
where $h(m)$ is $m$ modulo $p^{n-1}$. According to the isomorphism theorems,
$$M_{n-1}\approx M_n/G,\e$$
where $G=\ker(h)$. 
$G$ can be described as all the elements of $M_n$ which are equal to the 
unit matrix modulo $p^{n-1}$. Thus, inductively it is enogh to know $G$ 
and $M_{n-1}$.
$G$ is a group of $2^{m^2}$ elements and thus it is generated by the matrices 
$$M_{ij}^{ab}=\delta_{i,j}+p^{n-1}\delta_{a,i}\delta_{b,j},\e$$
where $a,b=1,2,\ldots,m$ and $i,j$ denote the entries of the matrix $M^{ab}$. The rest of $M_n$ is evidently generated by the
permutations and the element $A^{1,2}$. Counting orders it is easily established that these are indeed the generators modulo $G$: $A^{ab}$, $a\neq b$, along with, 
$$A_{ij}^b=\delta_{i,j}+p^b \delta_{i,a}\delta_{j,a},\e$$ 
where $b=1,2,\ldots n-2$,
are generators of $M_{n-1}$, which follows by induction on $n$, i.e., the $A$'s are generators 
of $M_{n-1}$ modulo $G$. Thus we count all the group elements $(p^{n-1}\times p)^{m^2}=p^{nm^2}=|M_n|$. 
Finally by acting with permutations, it is enough to have $M^{12}$ 
and $M^{p^{n-1}}$, arriving at theorem (1).

The proof of theorem (2) uses a similar filtration. Let $N$ be the rank 
$m$ matrices over $2^n$. Assume that the theorem holds for $n-1$, then 
by considering $B\in N$ modulo $2^{n-1}$, it follows that $B$ is equal 
up to a similarity transformation to one of the three classes, modulo 
$2^{n-1}$.
For the diagonal classes, we do not need to bother as the theorem was  proved earlier (it is the same as for any $p$).
Say, it is equal to 
$$\twidle N=\pmatrix{0&1&0&\ldots\cr1&0&0&\ldots\cr0&0&1&\ldots\cr\ldots&\ldots
\ldots&\ldots\cr}\,\mod 2^{n-1},\e$$
The $N$ is the same as $\twidle N$ up to some addition of $2^{n-1}$
in some entries. These can be easily eliminated by the similarity 
transformations $A^{a,b}$ (for the entry $(a,b)$) or $M^b$ for the entry
$(b,b)$. For type (II) the proof is the same. We now use theorem (1) to
establish that the three types are distinct, since we used all the generators. (Of course we can still freely permute the rows and columns of
the matrix, but this does not change the result.)

It, thus, remains to prove the therem for $n=1$. We use induction on $m$.
Say $N_{1,i}=1\mod 2$, then also $N_{i,1}=1\mod2$. We can use these two 
elements to "clean" the first and $i$'th rows and columns as above, except for these two 
elements. Thus we decompose off this matrix and the proof follows.

As an amusement the reader may wonder how theorem (1) works for rank $1$
matrices, i.e., numbers $m=0,1,2,\ldots,p^n-1$ modulo $p^n$, which are $1$ 
modulo $p$. Here we 
multiply $m$ by $p^{n-1}+1$ enough times to make $0<m<p^{n-1}$ (this is always possible since multiplying by it is like adding $p^{n-1}$).
Then we repeat these steps inductively. When we continue we might have to 
keep multiplying with $p^{n-1}+1$ for higher $n$ to compensate for the fact that we are working modulo a higher integer.
This presents any integer modulo $p^n$, which is $1\mod p$, as a product
of the generators $p^b+1$, $b=1,2,\ldots n-1$.

From theorem (2) it follows that,
we need to worry only, $p=2$, 
about rank two, and $n_i=n_j$.

Now, let us turn to the problem of realizing the non--regular automorphisms,
eq. (17). We employ the following alogarithm. First, from eq. (5) the central
charge is determined, $c$. We denote the lattice by $M$ and its matrix
of scalar products by $A$, $A_{ij}=\alpha_i\alpha_j$, where $\alpha_i$ are the
basis vectors for the lattice. We assume that $A$ is of the form,
$$A=\pmatrix{a&b&c&0&0&\ldots\cr b&d&e&0&0&\ldots\cr
c&e&f&-1&0&\ldots\cr 0&0&-1&2&-1&\ldots\cr\ldots&\ldots&\ldots&
\ldots&\ldots&\ldots\cr},\e$$
i.e., $A$ is the same as the Cartan matrix of $SU(n+1)$ except for a
replacement of a $3\times 3$ block at the top.

We run over all the possible matrices, $A$, where the coefficients are in a
certain range, and the diagonal elements
are positive integers (to describe an even lattice), and also
$A_{ij}^2 < A_{ii}A_{jj}$ to ensure that all the angles are real.

Below we quote the result of the computer run. To this 
avail we define the rank $n$ matrix,
$$U^{x,y,z,a,b,c}_n=\pmatrix{2x&-a&-b&0&0&\ldots\cr
                           -a&2y&-c&0&0&\ldots\cr
                           -b&-c&2z&-1&0&\ldots\cr
                            0&0&-1&2&-1&\ldots\cr
                            0&0&0&-1&2&\ldots\cr                          \ldots&\ldots&\ldots&\ldots&\ldots&\ldots\cr},\e$$
where the lower right hand side of the matrix is 
the Dynkin matrix of $SU(n)$.

We find for $Z_k\times Z_k$ with the non--diagonal automorphism and $k=2^m$:

1) $m$ odd: $M_1=U^{k/2,k,1,k,\sqrt{k/2},0}_8,$

2) $m$ even: $M_2=U^{1,k,{k+8\over12},0,1,{k\over2}}_8,$

(the simplest case $m=1$ also affords the Lie algebra realization $M=D_8$).
It can be checked that the determinant is $k^2$ as it should
and that the dimensions and orders are those corresponding to 
the non-diagonal automorphism of $Z_k\times Z_k$.
The characteristics of this are as follows:

1) All elements of $M^{-1}$ are of the form $n_{ij}/k$ where
$n_{ij}$ are integers. The diagonal elements $n_{i,i}$ are all even.

2) At least one $n_{ij}$ is odd, say $n_{\alpha,\beta}$

3) If both $n_{\alpha,\alpha}$ and $n_{\beta,\beta}$
are $2\mod4$ it is an automorphism of type (II). If not, it is
type (I).

We find that $M_1$ is always type (I), wheras $M_2$ is always
type (II) except for $k=4$ where it is of type (I). The matrix $h$ can be computed directly by $h_{\alpha\beta}=p M^{-1}_{\alpha\beta}$. 

To complete the realization, we find for even $m\geq4$ the type (I) realizations: 
$$M=U_8^{k/4,5k/16,5,k/4,\sqrt{k}/2,7\sqrt{k}/4},$$
and for $k=4$ an example of a type (II) realization is,
$$M=U_8^{4,5,2,4,4,1}.$$
For odd $m$ the type (II) realizations have $c=4\mod 8$ and we find,
$$M=U_4^{k,k,1,k,\sqrt{2k},0},$$
as a realization. 

This completes the realizations of all non regular automorphisms. 
The foregoing discussion along with the results of ref. \r\BGG\ prove the following theorem, previously conjectured in
ref \r\Found:

\Theorem: All pseudo bosonic systems can be realized by ordinary bosons. The actual realizations are given explicitly as ordinary bosons propagating on a rational even lattice with an extended algebra of integral dimensions.

Remark: Actually we realized every modular matrix, $S$. 
To realize every possible $T$ it is the same. 
We just need to replace
$|G|$ by 2$|G|$, i.e., $h$ is defined modulo $2$. The rest of
the proof is identical, as is the result. 

\refout\bye